\documentclass[useAMS,usenatbib]{mn2e}
\usepackage{natbib}
\usepackage{epsfig}
\usepackage{aas_macros}
\usepackage{color}
\usepackage{soul}
\newcommand{\removeR}[1]{}
\newcommand{\changeR}[1]{\textcolor{black}{#1}}

\title[The Earth transiting the Sun as seen from Jupiter: an inverse 
Rossitter-McLaughlin effect]{The  Earth transiting  the Sun as seen from 
Jupiter's moons: detection of  an inverse Rossiter-McLaughlin effect produced by the  
Opposition Surge  of  the icy Europa.}
\author[P. Molaro,  M. Barbieri, L. Monaco, S. Zaggia and C. Lovis]{P. 
Molaro$^{1}$\thanks{E-mail:
molaro@inaf.oats.it (PM)}, M. Barbieri$^{2}$, L.
Monaco$^{3}$ , S. Zaggia$^{4}$ and C. Lovis$^{5}$\footnotemark[1]\thanks{ Based 
on observations collected at the European Souther Observatory, Chile.  Program
 ESO  N.  092.C-0832(E) and at  the Italian Telescopio Nazionale Galileo (TNG) operated on the island of La Palma by the Fundaci—n Galileo Galilei of the INAF (Istituto Nazionale di Astrofisica) at the Spanish Observatorio del Roque de los Muchachos of the Instituto de Astrofisica de Canarias. Program A28 TAC-22
 }\\
$^{1}$ INAF-Osservatorio Astronomico di Trieste,  Via G.B. Tiepolo 11, Trieste I 
34143,  Italy\\
$^{2}$ Department of Physics, University of Atacama, Copayapu 485, Copiapo,  
Chile\\
$^{3}$ Departamento de Ciencias Fisicas, Universidad Andres Bello, 
Rep\'ublica 220, 837-0134 Santiago, Chile \\
$^{4}$ INAF-Osservatorio Astronomico di Padova, Vicolo dell'Osservatorio 5, 
35122, Padova, Italy\\
$^{5}$ Geneva Observatory, University of Geneva, Ch. des Maillettes 51, 1290, 
Versoix, Switzerland }
\begin{document}

\date{Accepted.... Received 2012}

\pagerange{\pageref{firstpage}--\pageref{lastpage}} \pubyear{2002}

\maketitle

\label{firstpage}

\begin{abstract}

We report on a multi-wavelength observational campaign which  followed the Earth's  transit on  the 
Sun  as seen from  Jupiter  on  5 Jan the  2014.      Simultaneous  
observations of Jupiter's moons Europa and Ganymede  obtained with HARPS  from La Silla, Chile, and HARPS-N from La 
Palma, Canary Islands,   were performed to  
   measure the Rossiter-McLaughlin effect due to the Earth's passage  using the same   technique  successfully adopted   for the  2012 Venus Transit  
\citep{mol13}.  The expected  modulation in  radial velocities was   of  
$\approx$ 20 $cm~s^{-1}$ but an anomalous  drift as large as $\approx$ 38   
$m~s^{-1}$, i.e. more than two orders of magnitude higher and  opposite in sign, 
was  detected instead.  The consistent behaviour  of the two spectrographs  
rules out  instrumental origin of the radial velocity drift and BiSON 
observations  rule out the possible dependence on the Sun's magnetic activity.  We  
suggest  that   this anomaly is  produced by  the Opposition Surge on  Europa's icy surface, which amplifies  the intensity of the solar radiation 
from a  portion of the solar surface  centered around the crossing   Earth which can then be observed as a   a sort of   inverse Rossiter-McLaughling effect. in fact, a  simplified model of this effect can explain in  detail most features of the observed radial velocity anomalies, namely  the extensions  before and after the transit,  the small 
differences between the two observatories  and  the presence of a secondary peak 
closer to  Earth passage. This  phenomenon,   observed here for the first 
time,   should be observed  every time similar Earth alignments occur  with  
rocky bodies without atmospheres. We predict it should 
be observed again  during the  next conjunction of Earth and Jupiter in 2026.     
\end{abstract}


\section{Introduction}

Transits of Venus and Mercury in front of the Sun are  major historical events  
but  also other transits can be seen in the solar system  from other planets each 
time  the heliocentric conjunctions take place near one of the nodes of their 
orbits, with the exception  of  the innermost   Mercury. In particular,     the 
Earth can  be seen transiting   in front of the Sun from other planets.  These 
are rare events which were \removeR{estimated} \changeR{predicted} in detail by \citet{mee89}. For instance,  
the Earth will be seen transiting the Sun  from Mars only in  2084. \removeR{From  
Jupiter  this  happened} \changeR{As seen from Jupiter,  a  transit took place}  on  5 \changeR{January  2014.}    Next passage will be   grazing 
and  will \removeR{be} \changeR{occur} in 2026.
 \removeR{In} \changeR{During} these  transits\changeR{,}  the integrated solar light can be recorded  as \changeR{it is} 
reflected by \changeR{the planets   from which  the Earth is seen transiting in front of the Sun,}   offering a surrogate   direct watch as we  
showed with the observation  of  the  Venus Transit of 6  June 2012 when  we  
followed the transit  as if it were seen from the Moon \citep{mol13}.

 \removeR{Therefore,} \changeR{We}  planned an observational campaign   to \removeR{exploit} \changeR{observe} the Earth\changeR{'s} passage 
in front of the Sun  that \removeR{ occurred} \changeR{took place}  in January 2014. One of the motivations for 
\removeR{the} \changeR{this} observational campaign  was the detection   of the  Rossiter-McLaughlin  
(RM) effect\changeR{ on which we report } \removeR{on which we report} in this work.
      The RM  is  a radial velocity drift \removeR{due to} \changeR{caused by} the distortion of  the stellar 
line profiles \changeR{due to the occultation of the rotating stellar disk by an intervening body.} \removeR{when the  crossing of a body occurs in front of a  star and it 
occults   a fraction of the surface  of the rotating stellar disk.}
The   effect  was first predicted by Holt (1893), and discovered by
\citet{sch11},  and later confirmed  by \citet{ros24} and \citet{mcl24} in the 
eclipsing binaries $\beta$
Lyrae and Algol, respectively. \citet{sch00} suggested that the transit of a 
planet could also be
detected in the line profile of high signal-to-noise stellar spectra of
rotating stars, and a Jupiter-like planet was  first  observed in HD
209458 by \citet{que00}  with an amplitude of $\pm$
30   $m~s^{-1}$. The detection of  the RM effect provides information on 
the planet radius,  the angle $\lambda$ between the sky projections of the orbital axis
and the stellar rotational axis. Since then about 90  other Jupiters have been
observed, often with  very tilted orbits \citep{fab09, tri10, bro12,alb12}.  
  The smallest  RM effect  detected is  due to the Venus Transit  in front 
of the Sun of 6 June 2012  by  \citet{mol13} who used the integrated sunlight 
as reflected by the Moon at night-time
to record about half  transit by means of the high precision HARPS
spectrograph at the 3.6m La Silla ESO telescope. The observations performed in 
correspondence of
the passage of Venus in front of the receding solar hemisphere  showed
that the planet eclipse of the solar disk  was able to produce a modulation in
the radial velocity with an  amplitude of  $\approx$ -1   $m~s^{-1}$. The radial 
velocity
change  is comparable to the solar jitter and is more than one order of
magnitude smaller than that of  extra-solar hot Jupiters.

The amplitude of the radial velocity
anomaly stemming from the transit is strongly dependent on the projected
radius of the eclipsing body and on  the component of the
star's rotational velocity along the line of sight \citep{oht05,gim06,gau07}. A 
transit across a star with high
 projected rotational velocity produces a  radial velocity signature larger than 
across a slow
rotator. The radial velocity  drift  $\Delta V_s$ is given by:

 \begin{equation}
 \Delta V_s =   \frac{k^2}{1- k^2}\cdot \Omega_s\cdot  \delta_p\cdot \sin I_s
   \end{equation}

where $ \Omega_s$ is the stellar angular velocity, $\delta_p$ is the  projected position
of the planet on the stellar surface $\delta_p = (X_p^2+ Z_p^2)^{1/2}$, \changeR{I$_s$} \removeR{I$_s$} is the inclination between the stellar spin and the
y-axis and $k$ = R$_p$/R$_s$ isthe ratio between the planet and
stellar radii \citep{oht05}.

 \removeR{In} \changeR{During}  the  Earth\changeR{'s} transit   of  5  January 2014 the projected size of  the 
Earth  \removeR{is} \changeR{was}  about 1.3 $  \times 10^{-4}$  of the solar disk\changeR{. Assuming} \removeR{and taking} a solar  
rotation velocity of $v\sin I_\odot$ = 1.6 $\pm$ 0.3   $km~s^{-1}$  
\citep{pav12}
  the expected RM effect is  of the order of $\pm$ 20    $cm~s^{-1}$. \removeR{ Also} \changeR{Furthermore,} our 
Moon is \changeR{also }transiting   \removeR{on} the solar surface  \removeR{together with Earth}    but with  a 
delay of about four  hours \changeR{with respect to the Earth}.   This  type of configuration should  be  quite 
common in transits of  extrasolar planets which   likely   have also their own moons. \changeR{. The}
 \removeR{and the} expected RM effect due to the Moon\changeR{'s}  occultation is  of only few  
$cm~s^{-1}$.

  \section{\removeR{The} Observations}
 
 \subsection{Timing of the transit}
\begin{figure}
\centering
 \includegraphics[width=6.0cm]{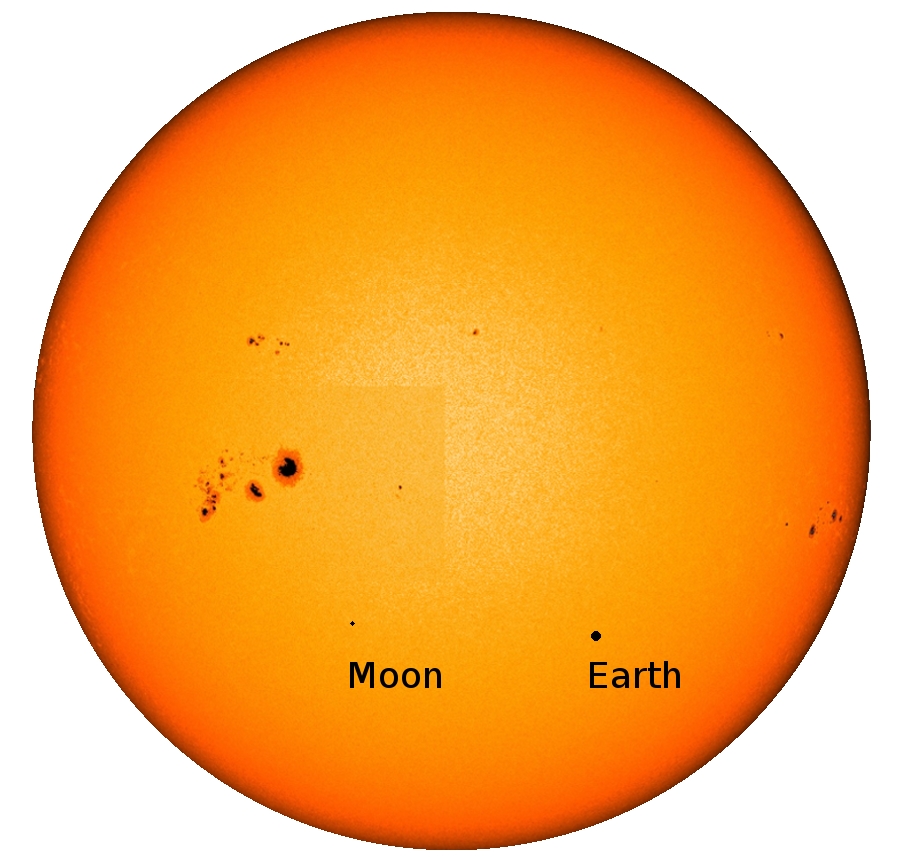}
 \caption{   Composite image of the Sun with Earth and Moon as seen from Europa at  19 UT of 5 January 2014. The sizes are in scale with the Earth\changeR{'s size} of 4.2  arcsec and the 
\removeR{whole} solar disk\changeR{'s size} of 369 arcsec. The solar image is from SDO/NASA HMI Intensitygram  at 617.3 $nm$ on 5 Jan 2014  and shows   prominent solar spots in the approaching 
solar hemisphere.
 }
 \label{figsole}
\end{figure}

 In Fig. \ref{figsole}   the Earth\changeR{,} \removeR{and} the Moon and their trajectories are shown 
  as they would appear to  an observer  on  Jupiter \changeR{(}or on  one of \removeR{his} \changeR{its} moons\changeR{)} \changeR{on  5 January 2014}.   
First contact \changeR{was} \removeR{is} at solar latitude of  -23.8$^o$ while the exit  \changeR{was} \removeR{is} at -35.7$^o$. 
  The heliographic latitude of the centre of the disk, the solar Bo angle\changeR{,}  \changeR{was} \removeR{is}   
of -3.6$^o$ and therefore the Sun was showing the south pole to  Jupiter with an 
inclination of 6$^o$ East of the solar axis. From  the Jovian system\changeR{,}   the  
black disk of the Earth  was of 4.2 arcsec  while 
 the    whole solar disk was of  369 arcsec\changeR{.} \removeR{   and  the} \changeR{The total} duration of the  passage 
was  of    9h40m.

\begin{figure}
\centering
 \includegraphics[width=8.0cm]{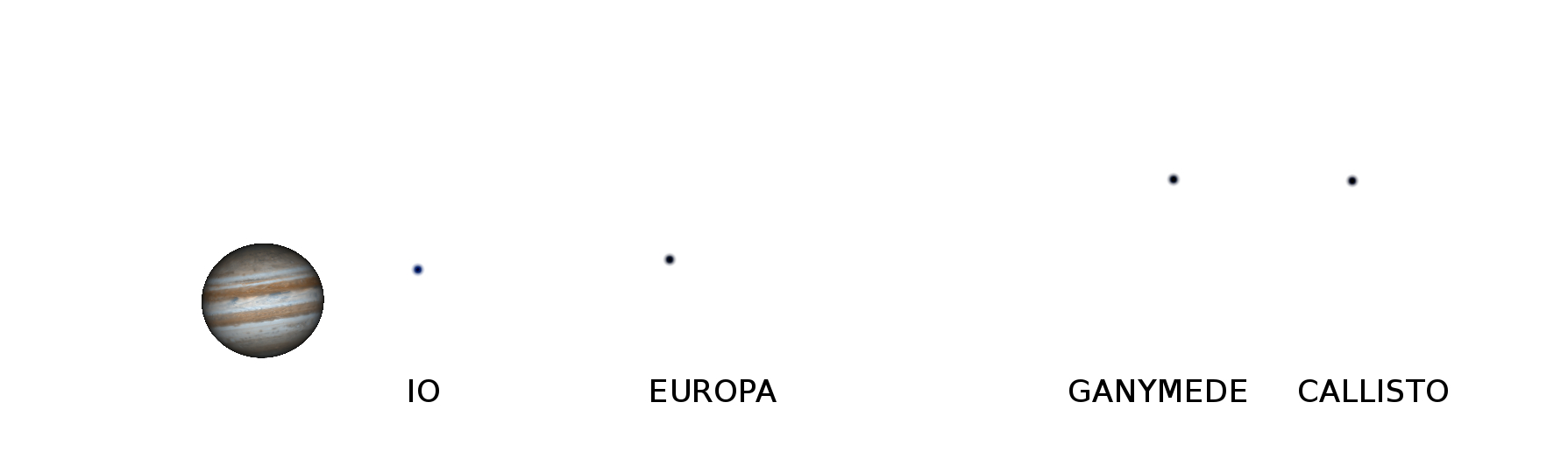}
\caption{ {\bf  Front view}  of the Jovian system on  5 January 2014 from an observer on the  
Sun or on the Earth.  From Jupiter the Earth transit    started at  MJD 
56662.70,  while  its moons arrive somewhat in  advance to the alignment, of about one  hour    for  Ganymede and      about  30 minutes 
 for  Europa, while  Io went behind Jupiter during the event.}
 \label{figgeometry}
\end{figure}

Jupiter itself is not a good sunlight reflector  due to its   high 
rotational velocity and to  the turbulent motions of its atmosphere.  Its major 
solid moons are better reflective mirrors. The geometrical configuration of the Jovian system is illustrated in  Fig. \ref{figgeometry}  from an observer on the Sun. The timing  of the Earth's transit  varies  
from one moon to another. In  January   the moons were seen  approaching  
Jupiter and therefore   arrived at the   alignment slightly before  the 
planet.
The Earth transit  on the reference frame of the jovian system started at  MJD 
56662.70  from Jupiter,  but it was seen    by     about  30 minutes 
in advance from Europa   and   about one  hour   from Ganymede.

 On January 2014 Jupiter could be seen at best from the northern hemisphere\changeR{,} but 
there was   not a suitable  site where  Jupiter could have been observed \removeR{along 
the   almost ten hours} \changeR{during the entire 10-hour transit}. 
 The moon   Europa was the best  suitable \removeR{replace of} \changeR{replacement for}  Jupiter\changeR{,} providing the most 
extended coverage  of the transit for about 6 hours from La Palma and  offering  
a limited possibility from La Silla    to follow for  $\approx$ 1 hour  the   
end  of the transit.  From La Silla \removeR{we could  start observing      at   } \changeR{it was possible to observe} the 
beginning  of  the night  when   Jupiter was rising  at    20 degrees over 
the horizon,  but remaining  always quite low and  reaching   35 deg  at the end of 
the Transit. 
The transit    could not   be observed from Mauna Kea either and   high resolution facilities that could   deliver very precise 
radial velocity measurements were not available in other 
astronomical sites.   Thus, La Palma and La Silla 
were the only  sites     where the phenomenon could  be  followed  
with  high resolution spectrographs suitable to deliver the required radial 
velocity precision.

\subsection{The observations}

The observations comprise a series of spectra taken with both HARPS-N and HARPS 
of the Jupiter's moons Europa and Ganymede
covering the range from 380 to 690 nm. At the epoch of the 
observations Europa and Ganymede  were fully illuminated and had a visual magnitude of 5.35  and 4.63 
mag and  apparent diameters of 1.02  and 1.72 arcsec, respectively. 
The   integration times of the observations were 60 or 120 s  and   delivered   a 
signal-to-noise ratio of $\approx$ 200 each at 550
nm with  a resolving power of R = $\lambda / \Delta\lambda$ $\approx$ 115000.  
The two  spectrographs at La Silla and La Palma are twins.  Both are  in 
vacuum, thermally isolated, stable and
equipped with an image scrambler which provides a uniform
spectrograph-pupil illumination which is essential   for  high  precision radial 
velocity observations.
HARPS was able to deliver a sequence of observations with a dispersion
of 0.64   $m~s^{-1}$  over a 500-day baseline for the radial velocity curve of  
an extra-solar planetary system composed by three
Neptune-mass planets \citep{lov06}.

\begin{figure}
\centering
  \includegraphics[width=65mm,angle= -90]{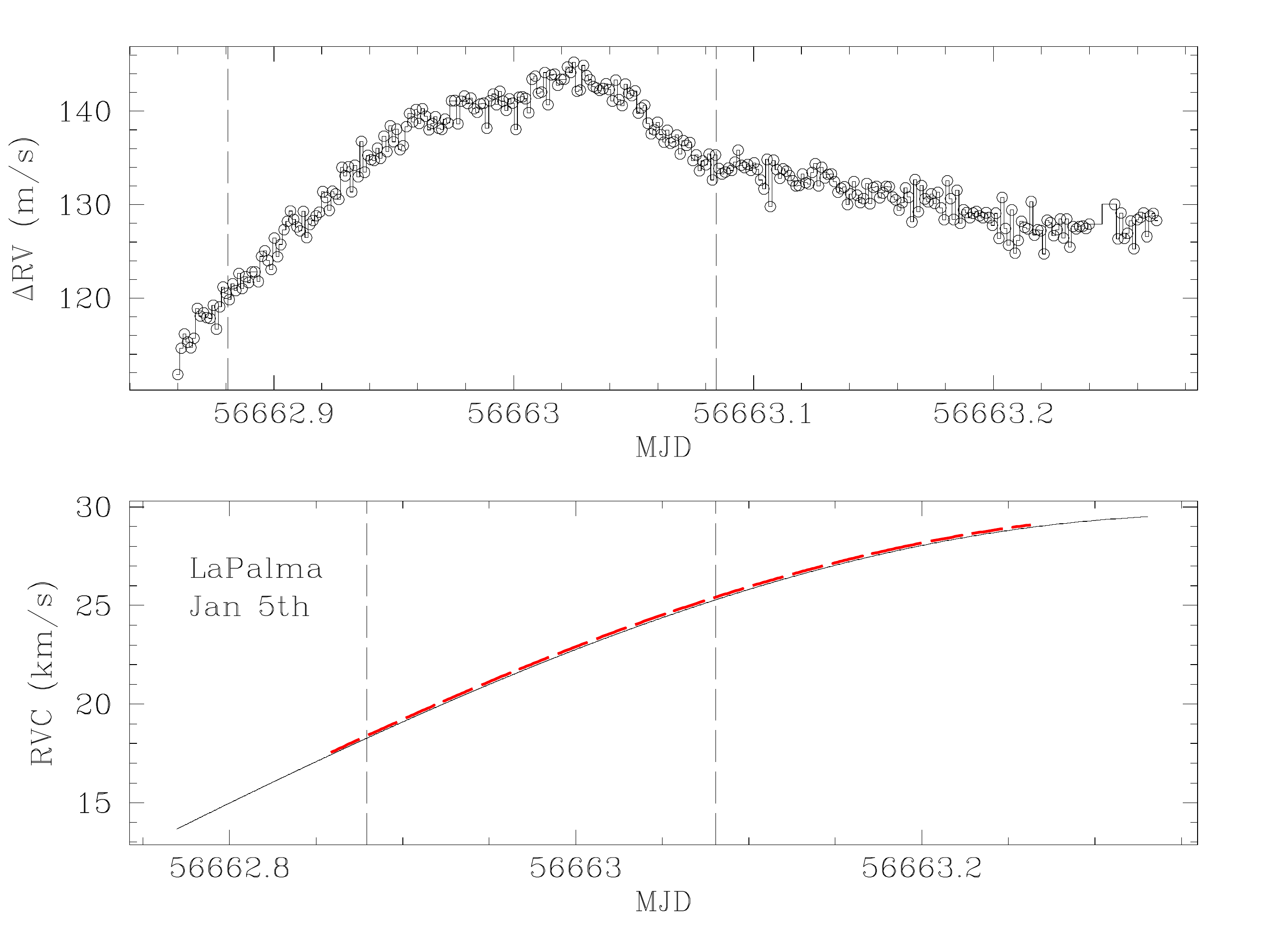}
 \caption{  Radial velocities measured  from  the   Europa  spectra of  5  
of January with HARPS-N observations.  The top  panel provides the RV corrected  
for the kinematical motions of the observer and of the moons\changeR{, which are} shown in the bottom 
panel with a black line. The red dotted line are the observed radial velocities.  \removeR{Note} \changeR{Notice} the 5 m oscillation of the Sun which is responsible \removeR{of} \changeR{for} \removeR{the} high 
frequency variation\changeR{s of} \removeR{with} an amplitude of $\pm$ 1  $m~s^{-1}$.  }
 \label{figlapalma2}
\end{figure}

The observations started  as soon as  Jupiter's  moons    became observable from 
the two sites.
We started    observing    Ganymede   from both telescopes on the 
night preceding  the transit to determine the pre-transit characteristic solar 
radial velocity.   At La Palma the observations  began   on   
2456661.983  MJD till   56662.265 MJD and   La Silla  
on    	56662.088  MJD till 56662.321 MJD.

The following night we  observed  Europa  from both telescopes to cover  the  second fraction  of the transit as much 
as possible.  At La Palma  observations  started    at MJD 56662.859 and   ended  at   MJD 56663.265,  while at La 
Silla observations were taken in the interval between     MJD 56663.070 and  
56663.330.

In the night following the transit we made   observations  of both   Europa and 
Ganymede      to determine the  post-transit characteristic solar radial 
velocity only  from La Silla. Observations of Europa   were taken from    
56664.068  MJD to   56664.167 MJD, followed by a sequence of observations of 
Ganymede till    56664.327 MJD.

\begin{figure*}
\centering
  \includegraphics[width=9cm,height=16cm, angle=-90]{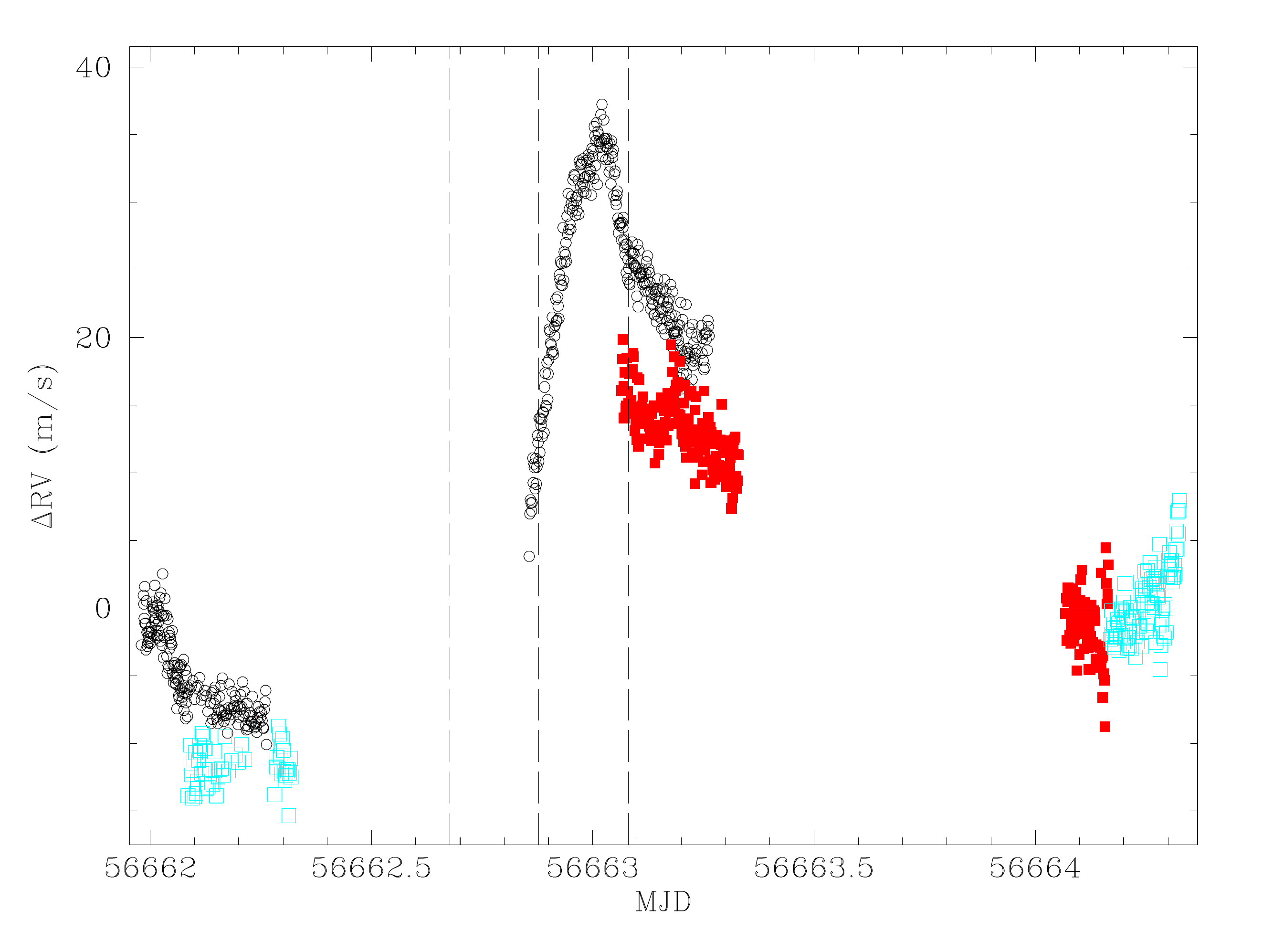}
 \caption{Radial velocities measured on  4-6 January 2014. Open black circles 
are observations of Europa from  La Palma  while color squares  are observations of 
Ganymede (cyan) and Europa (red) from La Silla. A constant  offset of  107.5   $m~s^{-1}$ as 
measured far out from the transit is taken as the instrumental baseline and is 
subtracted from the data. The vertical dashed lines mark the expected ingress and egress  of the Earth's transit as seen from Europa}
 \label{figobs}
\end{figure*}

\section{Radial velocities}
 
We used  HARPS and HARPS-N  pipelines to obtain the radial velocities from the 
observations.
The  pipelines  return a radial velocity  value  from the  cross--correlation of the 
spectrum with a G2 V flux template which is  the Fourier transform spectrometer 
(FTS) obtained by Kurucz at the McMath-Pierce Solar Telescope at Kitt Peak 
National Observatory \citep{kur84}.  The FTS solar spectrum  is calibrated onto 
the telluric emission lines and  is known 
to have  an  offset   in the zero point of the order of 100   $m~s^{-1}$
\citep{kur84,mol12}. 
The pipeline returns the radial velocity $RV_p$ relative to the solar system 
barycenter by taking the apparent position of the target. We thus subtracted the barycentric radial velocity correction, the {\it 
BERV},  which was  recorded in the fits headers to compute   the proper 
kinematical corrections.
  These included   the motions of the
observer  relative to  Jupiter's moons  at the instant when  the light received 
by the observer was reflected by the moons,   but also   the  radial velocity 
components of the motion of the moons relative to the Sun at the instant the 
light was emitted by the Sun  \citep{mol11, lan15b}. 
 The sunlight reflected by  Jupiter's  moons  is shifted by the heliocentric 
radial velocity of the moon with respect to the Sun at the time   the photons 
left  Jupiter's  moon and were shifted by the component of the Earth rotation 
towards the moon at the time  the photons reach  Earth. 
 The latter is the projection of the asteroid motion along the line-of-sight 
  adjusted for aberration, and comprises both 
the radial velocity of the moon  and the component from the Earth  rotation. 
 Thus, the  radial velocity   is:

 \begin{equation}
\centerline{$ RV = RV_{p} -  BERV   -   (RV_{moon-obs} + RV_{moon -  \sun })$.}	
 \end{equation}
  
 The quantities are computed by
using the JPL horizon ephemerides \footnote{Solar System Dynamics Group, 
Horizons Web Ephemerides Systems, JPL, Pasadena, CA 91109 USA 
http://ssd.jpl.nasa.gov}.
 The average rate  in the radial velocity change of Ganymede and Europa are   of 
 about 11  $m~s^{-1}$ and 12   $m~s^{-1}$  per minute, respectively. During the 
exposure of one or two minutes this velocity
  change produces some spectral smearing. However,  we apply the  corrections   
to mid-exposure values  and since the spectral smearing is symmetrical to a  
good approximation  it does not result into  a net shift in the measured radial 
velocities.

    Fig.   \ref{figlapalma2}   in the top panel shows  the     corrected radial 
velocities for the observations taken at La Palma on the 5th of January. These 
are  obtained from the radial velocities returned by HARPS pipeline once    
the kinematical corrections described  above, and shown in the bottom panel of 
the figure,  are applied.

 The values do  not show  clear discontinuities in connection with the Earth 
transit and suggest a  complex behaviour.
  It must be noted that there is a   known offset  in absolute radial velocities which  originates from the 
use of the FTS solar spectrum as a template.   This was  measured in 102.5  
$m~s^{-1}$ \citep{mol13}   in coincidence of the Venus transit with an   
 uncertainty of the order of few  $m~s^{-1}$ which  depends  from the   solar 
activity  of that day.

 Both spectrographs benefit of a 
 second fiber which supplies  ThAr spectra
simultaneous with    observations  and that  can be used to correct for
instrumental radial velocity drifts occurring over the night. The radial 
velocity differences
with respect to the previous calibration provide the instrumental drifts
for both spectrographs.

  \subsection{The Radial Velocity Anomaly}

The whole set of corrected solar radial velocities obtained from  the  Jupiter moon's 
spectra taken in the course of  the 3 nights from both sites    is shown in Fig.  
\ref{figobs} after subtraction of the radial velocity baseline. At the beginning 
of the observations  the RV is of 107 $m~s^{-1}$ while at the end it is at 108 
$m~s^{-1}$, and we adopt here a baseline of 107.5  $m~s^{-1}$ for simplicity.
The observations   taken at La Palma show a sudden drop    by about  7   
$m~s^{-1}$ after about one hour. Moreover, at the start of   La Silla sequence    the 
radial velocities were slightly lower   with  a difference   of about 4 $m~s^{-1}$  between the two 
spectrographs. 

In the following day   La Palma observations  started at about mid-transit   
with  radial velocities   rising very quickly  till they reached  a peak of  37 
$m~s^{-1}$. After the peak the radial velocities     declined  monotonically   
showing  a break in the slope in correspondence to  the end of the transit.   The vertical lines in 
the figure mark the  start, middle  and  end of the   transit for Europa. To note that the 
peak  of the  radial velocity is reached at MJD 56662.5 in correspondence of 3/4 of 
the Earth passage in the receding solar hemisphere  and the change in the slope in 
declining which corresponds to the end of the transit.  Both of them  will be 
discussed in the next section where we provide an interpretation of the 
phenomenon. 
 
In the night  following the transit  we made  observations only from La Silla. The
 radial velocities are  back to the values of the night preceding the transit.
  The observed pattern is completely at odd with   our expectations.
In the fraction  of transit covered by  observations  the Earth was eclipsing 
the receding solar hemisphere and the RM effect should have produced a  small
blue shift of the lines as a result of  the prevalence of light coming from the  
approaching solar hemisphere.  On the contrary we  observed  a change in the 
radial velocity of  37   $m~s^{-1}$  of opposite sign, i.e.  more than  two 
orders of magnitudes  greater than expected. Moreover, the  radial 
velocities did  not show any sharp change in correspondence of the end of the 
transit.
When the observations from the two spectrographs overlap in time  the radial 
velocity  behaviour  is similar in HARPS  and HARPS-N,  although there is a non 
negligible offset between the two   measurements.

\begin{figure}
\centering
 \includegraphics[width=8.6cm]{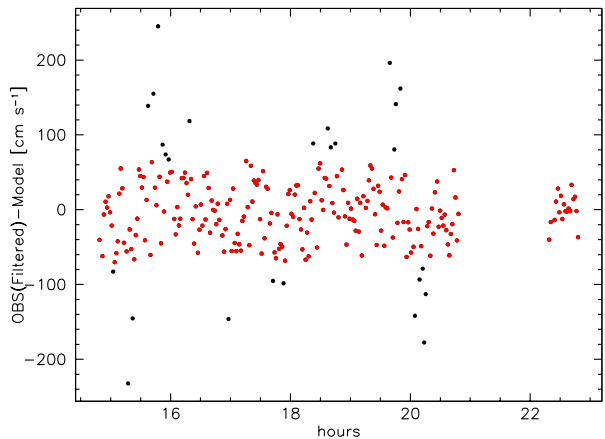}
 \caption{Residuals of the difference in units of  $cm~s^{-1}$ between the 
observed radial velocities and the RM model for the 2012 Venus transit described 
in \citet{mol13} but with the 5min oscillation filtered out. This illustrate the 
level of precision which can be achieved with HARPS at the sub $m~s^{-1}$  level 
in contrast with the radial velocity anomaly observed for the Earth's Transit 
which is more than  two orders of magnitude larger.}
 \label{figvenus2}
\end{figure}

The anomaly in radial velocity cannot have an instrumental origin. This is 
demonstrated by the fact that  the two observatories are giving consistent 
results and  similar RV anomalies have never been observed with 
HARPS .  An example of the precision which can be achieved in radial velocities 
with HARPS are the observations  of the Venus transit of 2012, which  were 
 taken  with the same technique adopted here.  For the Venus transit 
  we  obtained a remarkable   agreement between the predicted RM effect and  
observations. 
 In  Fig.  \ref{figvenus2} the difference between the RM model computed for the 
Venus passage  described in \citet{mol13} and the observations are plotted 
after the observations were  filtered for the 5m solar oscillations.   The 
residuals of the observations versus the model are of 0.55 $m~s^{-1}$, while the 
absolute difference is of -2 $cm~s^{-1}$, a difference which is within one sigma 
of the error in the  normalization of the observations with the radial 
velocities observed after the transit.
These observations were  treated   in the same way as those   we are dealing 
here showing  that large anomalies in radial velocities from HARPS are not  
plausible.   
Moreover, inspection of asteroid observations  taken with HARPS in its   life span 
of 12 years  show that RV  deviates from the mean by no more  than  $\approx$ 5 
$m~s^{-1}$ . Such deviations  are likely correlated with the solar magnetic activity as can be 
inferred from the presence of solar spots and  plages on the solar surface  
\citep{lan15}.

\begin{figure}
\centering
 \includegraphics[width=8.6cm, height=8cm]{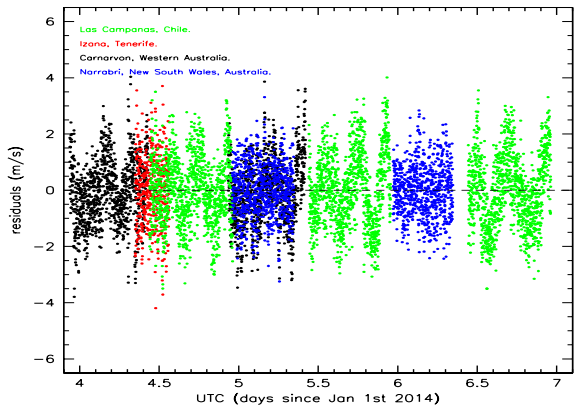}
 \caption{BiSON  solar observations archival data containing velocity residuals  
in the same days of our observations.  The data are  from Narrabri, New South 
Wales, Australia (red points),   Carnarvon, Western Australia,  (black), Izana, 
Tenerife  (red) and  Las Campanas, Chile (green). Courtesy of Steven Hale.}
 \label{fig_bison}
\end{figure}

Solar spots  could   also
affect the radial velocity of the solar lines  and  indeed in Fig \ref{figsole}  the solar
image  of   5   January   
 shows the presence of several solar
spots which could contribute at the level of few  $m~s^{-1}$. 
The  characteristic change
  is on a  time scale of solar rotation and  no effect is expected  during the relatively short  duration of   the Earth  transit.  The  
radial
velocity baseline  before and after  the transit also includes any contributions
originated by the presence of these solar spots.
  To check if   short-time strong solar  activity  occurred in coincidence of the 
transit  we  inspected the Birmingham Solar Oscillations Network (BiSON)   
archives containing solar velocity residuals  in the first days of January  
2014.  The data were captured from the sites of  Narrabri, New South Wales, 
Australia,  Carnarvon, Western Australia,  Izana, Tenerife and  Las Campanas, 
Chile and provide a continuos monitoring of the solar activity in proximity of 
the event.
The other  two  sites  of  Los Angeles and South Africa,   were 
offline in those days due to bad weather. The BiSON velocity residuals in Fig.  
\ref{fig_bison} do not show any anomaly at the level observed, and  suggest that 
the anomaly in RV we detected  does not depend from  an anomalous activity of the 
Sun.  In the next sections we will see that according to  our proposed 
explanation it is not a surprise that BiSON does not see the  radial 
velocity anomaly.

The effects of a microlensing onto the RM effect in the case of transiting 
planets has been studied in detail by \citet{osh13}.  The RM can vanish  in the  extreme 
cases of particularly massive planets, but it  has never been 
found to be inverted as we observed. Moreover, the  size of the Einstein ring  
due to  Earth observed  from Jupiter is  of only 47  $km$ which is   not 
expected to  produce any significant attenuation of the RM effect.

Therefore, we think  that this  effect is  real,  and we suggest   it   is due to 
the Opposition Surge  onto the icy Europa as we argue in detail in the next 
sections after a brief introduction to the nature of this  effect.

\subsection{The  Opposition Surge  effect }

The Opposition Surge  is a  
brightening of a rocky  celestial surface   when it
 is observed at opposition.  The  increase in brightness is a function of phase angle  and gets greater and greater   as  its phase angle of observation 
$\phi$  approaches zero.
The existence 
of the opposition surge was first recorded by \citet{geh56} but the precise physical origin  is not yet completely understood and  shadow hiding  or coherent backscatter have been proposed.

The former  stems from  the fact that when the light hits a 
rough surface at a small phase angles  all shadows decrease and the 
object is  illuminated by its largest extent.  
 It was Hugo von Seeliger  who  back in 1887 explained  the increase in albedo of  
Saturn's rings    to the corresponding  
reduction of the   shadows on the dust  particles of the rings    at opposition.    

 In the  coherent backscatter theory  the increase in 
brightness is due to a  constructive combination  of the light reflected  from 
the surface and by dust particles.   The  constructive  combination is achieved when 
  the size of the scatterers in the surface of the body is comparable to 
the wavelength of light.   At zero phase   
the  light paths will constructively interfere  resulting in an 
increase  of the intensity while as  the phase angle increases  the 
constructive interference decreases.   Coherent backscatter has been observed in 
radio  wavelengths and detailed physical models   are presented in 
\citet{hap93,hap02}  and \citet{shk01}.  

It is also possible  that  both  coherent backscatter and shadow 
hiding are operating. Which mechanism is dominant  depends  on the  physical 
properties of the surface such as porosity, the mean free path, and the single 
particle albedo.  Currently, theory is unable  to predict the amplitudes for 
either mechanism \citep{sch09}.  Considering both explanations
the  Opposition Surge is  also known as the Seeliger-Hapke effect.

\begin{figure*}
\centering
 \includegraphics[width=14cm]{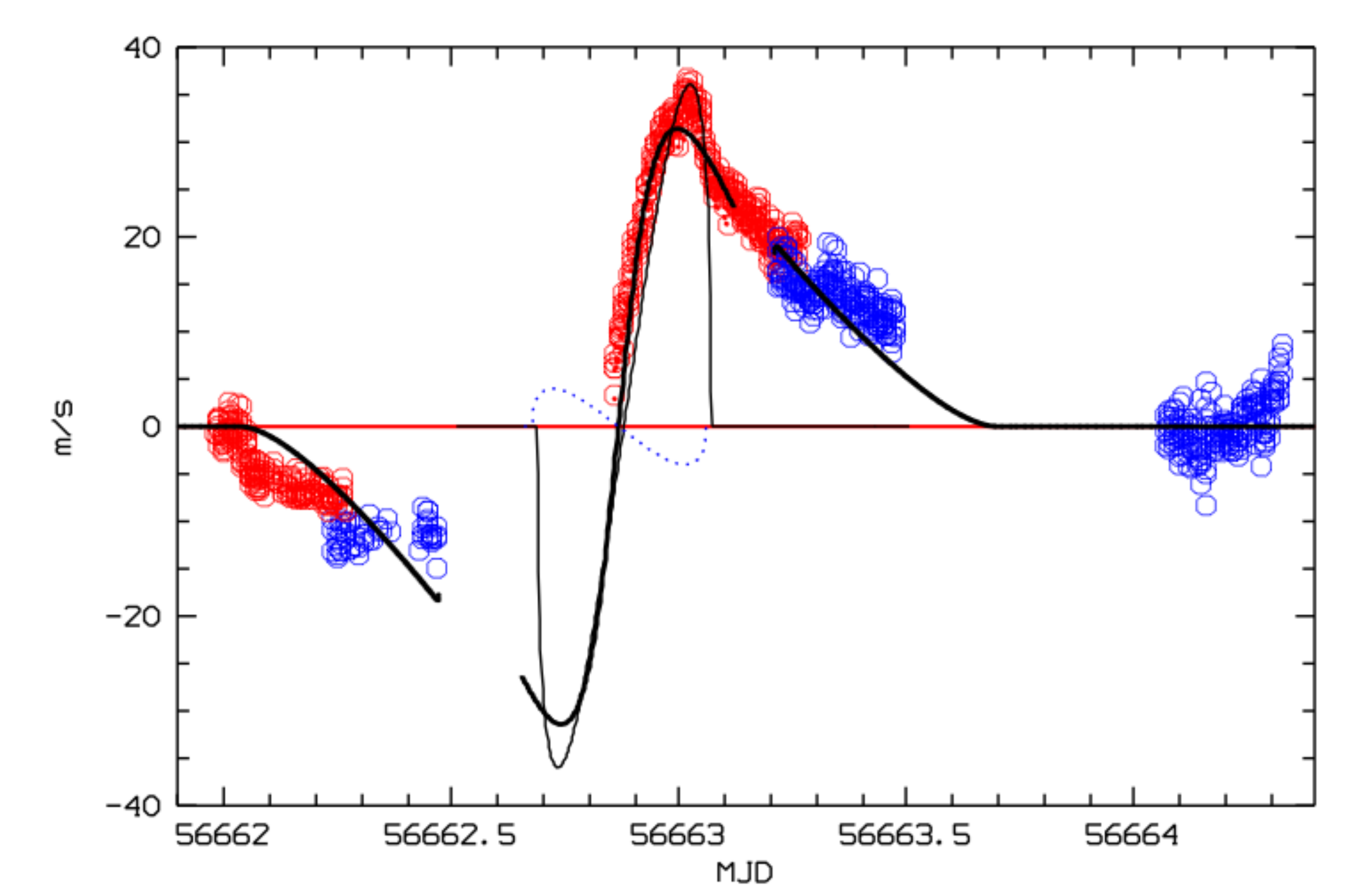}
 \caption{Model of the inverse Rossiter-McLaughlin RV  drift induced by an   increase in the solar 
emissivity  in the region behind the Earth trajectory due to the Opposition 
Surge. The thin black line shows the the drift expected from a small area moving with 
the Earth, while the thick line shows a drift coming from a larger area with 
radius of  $\approx$  10  arcminutes  required to match the start of the radial 
velocity drift.   The gap between the two models is due to the numerical 
impossibility to compute a RM effect for total  eclipse.
 The data points from la Silla (blue points) are delayed by 0.1468 MJD to 
compensate for the longitude difference between the observatories of La Palma 
and La Silla. Cfr text for the details.  The  blue dotted line  shows how the  expected Rossiter-McLaughlin  effect  but amplified by a factor of 50 since the real one  would have the thickness of the line. 
The inverse RM effect detected is about 400 times larger than the expected RM due to the Earth transit. }
 \label{figmodel}
\end{figure*}

\subsection{An Inverse Rossitter-McLaughlin effect}

In the following we argue  that  the Opposition Surge can explain   the radial 
velocity anomaly observed in  proximity  of the Earth Transit.
A characteristic feature of the Opposition Surge is the brightening of the 
planet  as  the phase angle $\phi$ decreases.  Solar  
photons which graze  the Earth have  smaller  angles than photons coming from 
regions of the solar disk far  away from the Earth edge.   Thus  they produce     
an   effective increase in the radiation coming from the region of the Sun just 
behind the Earth as it moves across the face of the Sun.   Along its 
passage  the Earth acts as a lens and the light   magnification     produces  a radial 
velocity drift which is  opposite in sign to that expected from a 
Rossiter-McLaughling effect, but of  identical  physical origin.  
The enhancement of a portion of the solar disk  produces  a  distortion in the 
solar line profiles with an asymmetric contribution from the two solar 
hemispheres of the same kind of the   Rossiter-McLaughlin. The opposite sign is 
because  instead of  an occultation  there is  an enhancement of the emission 
in  a restricted area of the solar surface.
 Instead of receiving less radiation from the hemisphere the Earth is crossing,  
due to its  occultation of the solar disk we are receiving more radiation from 
it  because of  the   enhancement  produced by the Opposition Surge effect of  the reflecting 
body. 
This  effect   not only compensates   the  effect of the partial solar eclipse 
by   Earth but   is able to produce  an opposite  radial 
velocity drift  by orders of magnitude stronger. 

Opposition Surge has been observed in  Jupiter's moon Europa and has become  
prominent for phase angles less than  $\phi$ $<$ 1 degree  \citep{sim04}. The 
Jovian moon has a  comparatively young surface    rich in water ice which 
produces a high  albedo. In these conditions coherent backscatter  
is expected to dominate over 
shadow hiding. However,   near infrared Cassini observations have been 
interpreted as   the Opposition Surge cannot be produced by coherent 
backscatter  alone,  but that it must have a significant shadow hiding  
component even in the presence of high albedo \citep{sim04}.

The Opposition Surge is not  fully understood and we cannot make a quantitative 
prediction of the  distribution of the light enhancement as a function of the 
angular distance from the Earth position.  
 
However, a simplified model  which accounts   for the  asymmetrical   emission 
from the two rotating solar hemispheres
can  explain most of the features  of the RV curve we observed. 

We  considered an   area around the Earth  with uniform enhanced emission  and 
we  computed the effect in RV as if it were  due to  the  RM effect. The sign  of the radial velocity drift  is reversed to simulate the emission instead of the eclipse.  The 
theoretical radial velocity anomaly of the Sun during the  transit is computed 
using the formalism  of \citet{gim06}. Since there is a degeneracy between    
the area and  the intensity of emission,   we just  scaled  the radial velocity 
to the observed one but preserved   the shape. The scaling factor   provides the 
amount of light enhancement which is necessary  for a given  area assuming a 
uniform emission, while    it is very likely that   it   changes within the area 
 as a function of  the phase angle. 

 We assumed the  rotational velocity  of the Sun  $V_{rot}$  is:
\begin{equation}
\omega = a + b\sin^2\phi + c\sin^4\phi
\end{equation}
where $\omega$ is the solar angular velocity measured in $^\circ$/day, 
$\phi$ is the solar latitude, and $a,b,c$ are the coefficients derived 
from the magnetic field pattern ($a=14.37, b=-2.3, c=-1.62$). The 
corresponding rotational velocity at latitude $\phi$ defined by the Earth  
trajectory is:
\begin{equation}
V_{\rm rot}=2\pi R_\odot\cdot(a + b\sin^2\phi + c\sin^4\phi)
\end{equation}
 
The limb darkening coefficients of the Sun    are $u_a=0.5524$ and 
$u_b=0.3637$,  taken 
from the tables of \citet{cla04}, for the $g$ filter and an Atlas model for the 
Sun with  solar metallicity, $T_{\rm eff}=5750$ K, $logg=4.5$, 
$\xi=1$  $km~s^{-1}$.

The  theoretical variation of the solar radial velocity 
during the transit computed with   the above derived  parameters is plotted as a 
thin line in Fig. \ref{figmodel}.
In the figure  it is possible to see that the radial velocity anomaly   does not 
end abruptly with the end of the Earth transit,   but it extends further after it.
    This is not surprising since  the  Opposition Surge does not last for the 
time of  eclipsing transit but it is also  present  when the Earth has just left 
 or is approaching  the solar disk,   
provided    the  solar rays are coming from  portions of the solar disk  which 
are at angles small enough   to produce the Opposition Surge. For many hours after the end of the Transit, the  Opposition 
Surge makes the solar hemisphere just left by  the Earth     brighter than the most 
distant one. Thus,  the radial 
velocity is  decreasing smoothly while  the Earth  is moving away and   the 
phase angle is increasing.

 We observe  the phenomenon extending after  the transit  but  not  its  end 
since  RV is still high   about six hours after the end of the transit.  It is 
only on the following night that we measured again a constant radial velocity.  
For symmetry  we can assume that the Opposition Surge   should also have started 
 many hours before the formal start of the Earth transit in coincidence with   
the sudden drop in radial velocities  by about 7   $m~s^{-1}$ observed on 
4  January  at MJD 5666.204-5666.206    from La Palma observations  of both 
Europa and Ganymede. Thus,  the Opposition Surge effect started to be effective 
and produced  a  radial velocity change  at  something  about   15 hours   
before the start of the Earth transit when  the Earth was  at a projected 
distance of  about 7 arcminutes from the solar edge.   We emphasize  that it is 
only the difference between the contributions of light coming from the two solar 
hemispheres that  matters.  An Opposition Surge  which provides equal enhancement of 
the two solar  hemispheres would  produce a brightening but not any detectable 
radial velocity change.  For symmetry the radial velocity anomaly  should have 
ended also 15 hours after the end of the  transit, i.e.  in a  period  which  is  
not  covered by our observations.

 In our simplified model   we have considered     a  circular region centered on 
the Earth and  a radius of 6 arcminutes, namely 165 times the projected radius 
of the Earth as seen from Europa. In Fig. 
\ref{figmodel} the computed radial velocities 
 are  overplotted to the observations approximately covering the transit   after scaling down the RM intensity by a factor of 30. The 
predicted radial velocity rise  follows the observations  quite well, though   it 
is somewhat less steep. The  peak    is reached when the Earth is approximately 
at about 3/4 of the solar receding hemisphere. This is  the position  where we expect the stronger effect on radial velocity due  to the combined 
effect of the almost tangential rotational velocity and of the limb darkening of 
the Sun.   During the decline  a break in the slope  with a more gentle decline is observed in proximity of the  Earth  
egress.  The region with enhanced emission has been 
enlarged to 10 arcminutes to  allow the RV anomaly to extend well  outside the 
transit. 
  
The first half of the transit could not be observed  either from la Palma or  La 
Silla and the observations cannot track   the passage of the  Earth  in front of 
the approaching solar hemisphere where the Opposition Surge should   have produced a  symmetrically negative  radial velocity behaviour. 
    Simultaneous observations  from the two observatories give slightly different 
radial velocities. Those  from la Silla are always lower  than those from la 
Palma (see Fig. \ref{figobs}). The difference is of about 4 $m~s^{-1}$  in the 
first night,  and of   about 10 $m~s^{-1}$  at the beginning of the second night,
but  they slightly decrease    to few  $m~s^{-1}$ as the event faded away.  While 
we cannot exclude some systematic offset between the two telescopes at the level 
of few  $m~s^{-1}$ the difference observed during the opposition seems a bit too 
large to be explained  only with this  systematic.  Thus, it  is quite possible 
that the  different locations on Earth of the two observatories  do not see 
exactly the same  Opposition Surge.  In particular, the distance from the Earth 
edge could have been relevant in determining the Opposition Surge intensity and 
therefore the radial velocity value. The  difference between the  longitudes  of La Palma  ( 28$^o$ 42.89' N, 17$^o$ 
54.29' W ) and La Silla (-29 15.67' S,  70$^o$ 43.88' W ) is of   0.1468 MJD, while the  distances 
from the equator is very similar.  This means that  after a time of  0.1468 MJD La Silla  will 
be at  approximately the same distance from the Earth edge  as La Palma.  In Fig. 
\ref{figmodel} we have  shifted   the data points from La 
Silla by this time difference and  they provide a much better  continuity and overlap with the values 
measured at La Palma.  To note that  this is achieved regardless of the fact that  
alignment of the Earth and of the Jovian systems has slightly changed in the 
meantime. This would  
 imply that    the intensity of the Opposition Surge is  very sensitive to the 
phase angle and therefore to the  location of the observer on Earth, in 
particular to its distance to the Earth's projected edges.

It is interesting to   note the possible presence of a double  peak   in 
proximity of the maximum of the radial velocity,   which is suggestive of  the 
presence of two components. While the broad  one could be  associated  with  a 
diffuse area  of  enhanced emission as we have discussed above, the latter  narrower 
one could be due to a stronger  emission located  in proximity of the Earth. 
The result  of an emission from a   relatively small  area  in proximity of the  
Earth  is plotted in Fig. \ref{figmodel} as a thin line  which   reproduces 
quite well the peak with a small delay  of + 0.01 MJD.

As we noted above  no  radial velocity  anomalies were observed   during 
the Venus Transit of  6  June 2012.     The  Moon was   in opposition at  
about
8 degrees ahead of the Earth  at a  phase  angle large enough to avoid the  
Opposition Surge. \citet{yok99, bur96}  with  their study of Clementine data 
estimated a 30-40$\%$ increase in brightness of the Moon  when  going from 4 to 0 degrees of phase 
angle.  
However, even if present, this should   not have produced a radial velocity anomaly 
since it would not have been  connected to the Venus passage in front of the 
Sun. For  similar reasons the BiSON measurements obtained from a direct watch of 
the Sun  do not see the radial velocity anomaly which is  produced by the   
magnification of a portion of the Sun    induced by the Opposition Surge on the 
Earth passage.

We note also that the presence of a strong Opposition Surge during the Earth 
transit   is probably the explanation  of the  lack of 
detection of  the luminosity drop in the flux due to the Earth occultation  
which has been searched for unsuccessfully by many teams.

\section{Conclusions}

  We followed the Earth transit of  5  Jan 2014    as seen from  Jupiter by 
means of    observations of   Jupiter's moons Europa and Ganymede. 
  The observations were made   with  HARPS
spectrograph at  La Silla, Chile, and with  HARPS-N spectrographs from La Palma, 
Canary Islands, originally aimed     to detect    the RM effect due to the Earth 
passage on the face of the Sun. We  followed   the   same    technique successfully 
adopted   for the  2012 Venus Transit  \citep{mol13} where the  RM effect  was 
measured and found in  agreement with the theoretical model within few  
$cm~s^{-1}$. In the case of the Earth transit  the expected  modulation in  
radial velocities was   of  $\approx$ 20  $cm~s^{-1}$.
  Instead,   an anomalous  and very large radial velocity  drift was observed. 
The half amplitude of the radial velocity drift observed was  as large as 35  $m~s^{-1}$, i.e. about  four hundred times  higher and  opposite in sign.
  
   The   similar behaviour  in the observations taken from both 
telescopes rules out an instrumental origin and  suggests a  physical origin
    which we identified   as the product of the   Opposition Surge   effect onto 
 Europa's icy surface. The Opposition Surge  effect amplifies  the intensity of 
the solar radiation from the portion of the Sun crossed by the Earth  and produces  a sort of   inverse Rossiter-McLaughling. This phenomenon  has never 
been observed before and is associated to the rather unique geometry in which we 
observed the Earth transit. In fact, simultaneous radial velocities obtained by 
BiSON through a direct solar watch  do not show the radial velocity anomaly and rule out 
 that they originate in the Sun.

A toy  model which assumes an enhancement of the solar radiation from a  
projected solar region centered on the  Earth's position   produced by the 
Opposition Surge  explains  the general behaviour shown by the radial velocity 
measurements. In particular,  we are able to explain why the anomaly  is observed  
 also   before and after  the Earth Transit, and the differences in radial 
velocities measured by  the two observatories as due to the different distances 
from the Earth edge,  as well as the presence of  a second peak associated  with the 
smaller projected solar region around the Earth but with greater intensity,  and 
why   we did not see a similar anomaly  in the 2012 Venus Transit.  

The Opposition Surge effect   provides a coherent and  plausible description of 
the  anomaly in radial velocity  as an inverse Rossitter-McLaughlin  that we  
observed for the first time during the Earth Transit.
 The effect  could be observed again  every time   the Earth is seen in transit 
against the Sun from other planets or smaller bodies  in the solar system.
  
  The next  Earth transit will   occur from jupiter     in 2026,    but it will be a  
grazing transit  quite unfavourable  to any kind of observations \citep{mee89}.
However, since we have observed the effect of the Opposition Surge when the 
Earth was at  an angle as high as   about 10 arcminutes,  we can predict that this 
same phenomenon can be observed again although with a minor amplitude in radial 
velocities.

\section*{Acknowledgments}

We warmly thank Steven Hale for providing the BiSON data of the days of  our 
observations. Very useful discussions with Emilio
Molinari, Gaspare Lo Curto, Claudio Lopresti and  Gerardo Avila in different 
stages of this work are also
acknowledged.   We thank also Harutyunyan Avet for his competent assistance with 
the HARPS-N observations.

\bsp

\label{lastpage}

 \bibliography{transito_biblio}
 
\end{document}